\documentclass[letters, usenatbib, fleqn, a4paper]{mnras}

\usepackage{newtxtext, newtxmath}
\usepackage[T1]{fontenc}
\usepackage{ae, aecompl, times}

\usepackage{graphicx}
\usepackage{amsmath}
\usepackage{amssymb}
\usepackage[math]{cellspace}

\usepackage{pstricks}

\usepackage{soul}

\usepackage{aas_macros}
\usepackage{bm}

\title[On the dust-to-gas ratio in molecular clouds]{Is the dust-to-gas ratio constant in molecular clouds?}

\author[Tricco, Price \& Laibe]{Terrence S. Tricco$^{1}$\thanks{E-mail: \href{mailto:ttricco@cita.utoronto.ca}{ttricco@cita.utoronto.ca}},
Daniel J. Price${^2}$
and Guillaume Laibe${^3}$ \\
$^1$ Canadian Institute for Theoretical Astrophysics, University of Toronto, 60 St. George Street, Toronto, ON M5S 3H8, Canada \\
$^2$ Monash Centre for Astrophysics and School of Physics \& Astronomy, Monash University, Clayton, VIC 3800, Australia \\
$^3$ Univ Lyon, Univ Lyon1, Ens de Lyon, CNRS, Centre de Recherche Astrophysique de Lyon UMR5574, F-69230, Saint-Genis-Laval, France}

\date{\today}

\pubyear{2017}

\begin{document}
\label{firstpage}
\pagerange{\pageref{firstpage}--\pageref{lastpage}}

\maketitle

\begin{abstract}
We perform numerical simulations of dusty, supersonic turbulence in molecular clouds. We model 0.1, 1 and 10 $\mu$m sized dust grains at an initial dust-to-gas mass ratio of 1:100, solving the equations of combined gas and dust dynamics where the dust is coupled to the gas through a drag term. We show that, for 0.1 and 1 $\mu$m grains, the dust-to-gas ratio deviates by typically 10--20\% from the mean, since the stopping time of the dust due to gas drag is short compared to the dynamical time. Contrary to previous findings, we find no evidence for orders of magnitude fluctuation in the dust-to-gas ratio for 0.1 $\mu$m grains. Larger, 10 $\mu$m dust grains may have dust-to-gas ratios increased by up to an order of magnitude locally. Both small (0.1 $\mu$m) and large ($\gtrsim 1$ $\mu$m) grains trace the large-scale morphology of the gas, however we find evidence for `size-sorting' of grains, where turbulence preferentially concentrates larger grains into dense regions. Size-sorting may help to explain observations of `coreshine' from dark clouds, and why extinction laws differ along lines of sight through molecular clouds in the Milky Way compared to the diffuse interstellar medium. 
\end{abstract}






\section{Introduction}

 The ratio of dust to gas mass in the Milky Way is long established to be around 1:100 in the diffuse interstellar medium (ISM), with 1\% dust compared to gas \citep{bsd78}. This ISM value is commonly adopted to infer the mass of  molecular clouds from extinction mapping \citep[e.g.][]{lombardietal14}, but changes in dust properties can have dramatic consequences for inferred cloud masses (for example, the recalibration adopted by \citealt{evansetal09} resulted in a 40\% change in the resulting cloud mass estimates).
 
Whether or not the ISM dust-to-gas ratio applies within molecular clouds is an open question. Lines of sight through molecular clouds are known to have anomalous extinction laws, best fit by models with $R_{V} = A_{V} /E(B-V)\approx 5$ instead of the more typical $R_{V}=3.1$ in the diffuse ISM \citep*{ccm89,wd01}. This implies a change in the grain size distribution from the typical \citet*{mrn77} power-law \citep*{kmh94}, and is usually attributed to grain growth \citep[e.g.][]{chapmanetal09}. The distribution of grain sizes in the Milky Way is peaked at a radius of $\sim$0.1 $\mu$m \citep{wd01, draine03}, and the presence of micron and larger sized grains is controversial. There are few constraints on the abundances of large grains, but their presence is thought to explain the `core shine' effect seen in mid- and near-infrared observations of dark clouds \citep{paganietal10, steinackeretal10, lefevreetal14}. There also exists observational evidence for local variations of the dust-to-gas ratio within molecular clouds. \citet{liseauetal15} examined dust-to-gas ratios across the $\rho$ Oph A molecular cloud core, using N$_2$H$^+$ emission as a gas tracer, finding a mean dust-to-gas ratio of $\sim 1.1$\%, not far from the canonical value, but with localised values ranging from 0.5\% to up to 10\%.

An increase in the mean grain size in dense gas may also result from grain dynamics. \citet{padoanetal06} found that the power spectrum of near infrared extinction maps in Taurus was significantly shallower than the power spectrum of the corresponding $^{13}$CO map, suggesting intrinsic fluctuations in the dust-to-gas ratio caused by dynamical decoupling of gas and dust. Preliminary simulations reported by \citet{padoanetal06} found that turbulence could generate significant small-scale fluctuations in the dust-to-gas ratio. More recently, \citet{hl16} and \citet*{lhs17} performed simulations of $0.1$~$\mu$m dust in molecular clouds, finding that the dust-to-gas ratio could `exhibit dramatic fluctuations' (orders of magnitude), with dust filaments appearing even in the absence of gas filaments, leading to the possibility of `totally metal' stars formed in regions of extreme metallicity concentration \citep{hopkins14}.

The problem with the preceding numerical studies is that they used tracer particles to model molecular cloud dust. \citet{pf10} showed that tracer particles in simulations of supersonic turbulence do not accurately capture the dynamics, producing numerical artefacts in the form of exaggerated concentration in high density regions and almost total absence in underdense regions. Furthermore, for two fluid dust and gas mixtures at high drag (small grains), \citet{lp12a} proved that it is necessary that the gas resolve the `stopping length' of the grains, $l \sim c_{\rm s} t_{\rm s}$,  to correctly predict the dust dynamics (where $c_{\rm s}$ is the sound speed and $t_{\rm s}$ the dust stopping time). For 0.1 $\mu$m dust grains in a 10 pc sized molecular cloud such as in the simulations of \citet{hl16}, this would require $6400^3$ gas resolution elements. If this spatial resolution requirement is not satisfied (as it was not in their paper), then spuriously high dust concentrations are produced as dust particles become trapped on scales below the gas resolution length.

In this Letter, we investigate dynamical variations of the dust-to-gas ratio in molecular clouds caused by the finite stopping time of the dust grains using three-dimensional numerical calculations of dust-gas mixtures in non-self-gravitating, turbulent molecular clouds. Importantly, we use the single fluid dust/gas model of \citet{lp14a, lp14b} and \citet{pl15}, which avoids the spatial resolution requirement of dust tracer particles or a two fluid method. The one fluid equations and our numerical method are described in Section~\ref{sec:details}. Simulation results are presented in Section~\ref{sec:results} and discussed in Section~\ref{sec:discussion}. We summarise in Section~\ref{sec:summary}.


\section{Simulation details}
\label{sec:details}

\begin{figure*}
\centering
\setlength{\tabcolsep}{0.002\linewidth}
\begin{tabular}{cccl}
\includegraphics[width=0.28\linewidth]{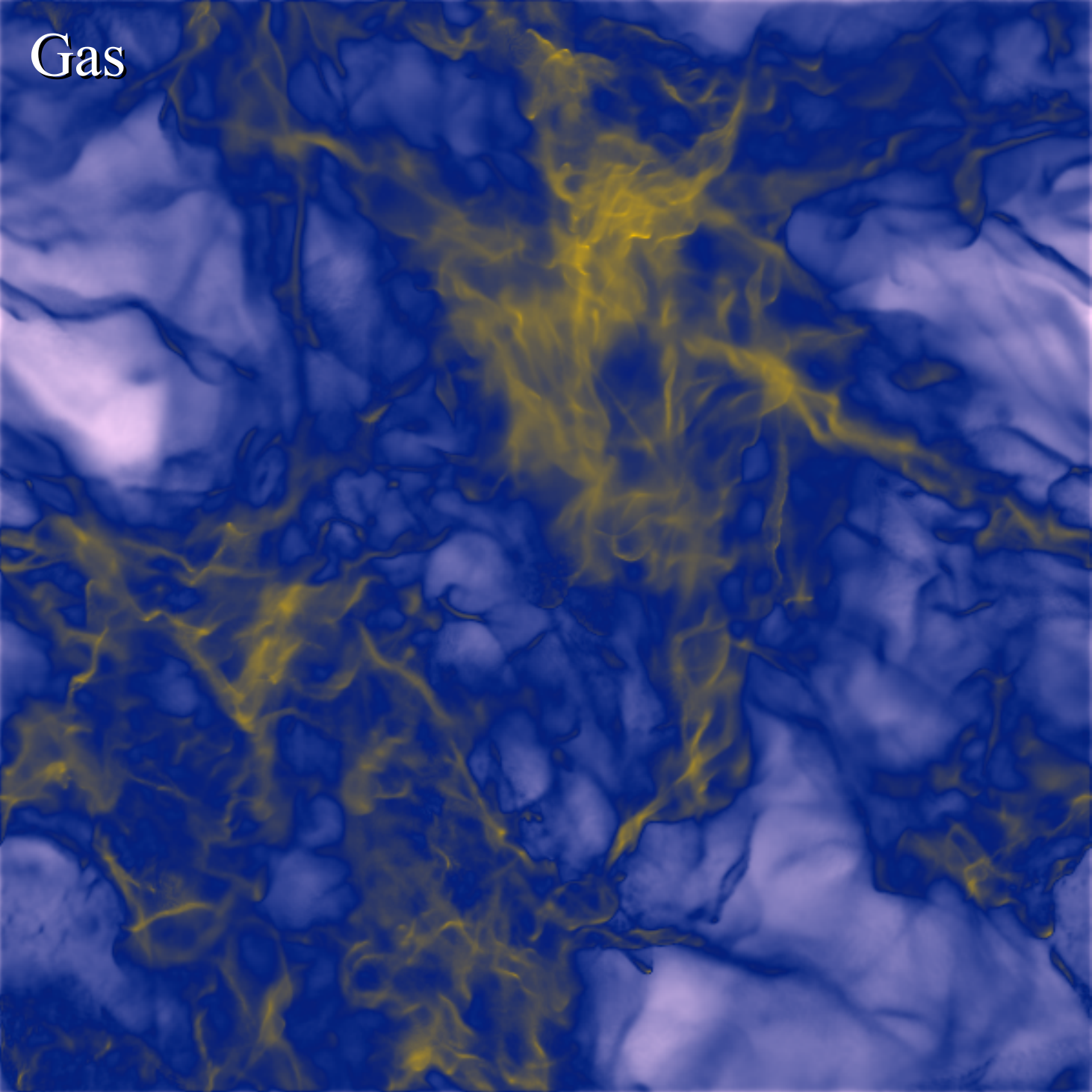} 
 & \includegraphics[width=0.28\linewidth]{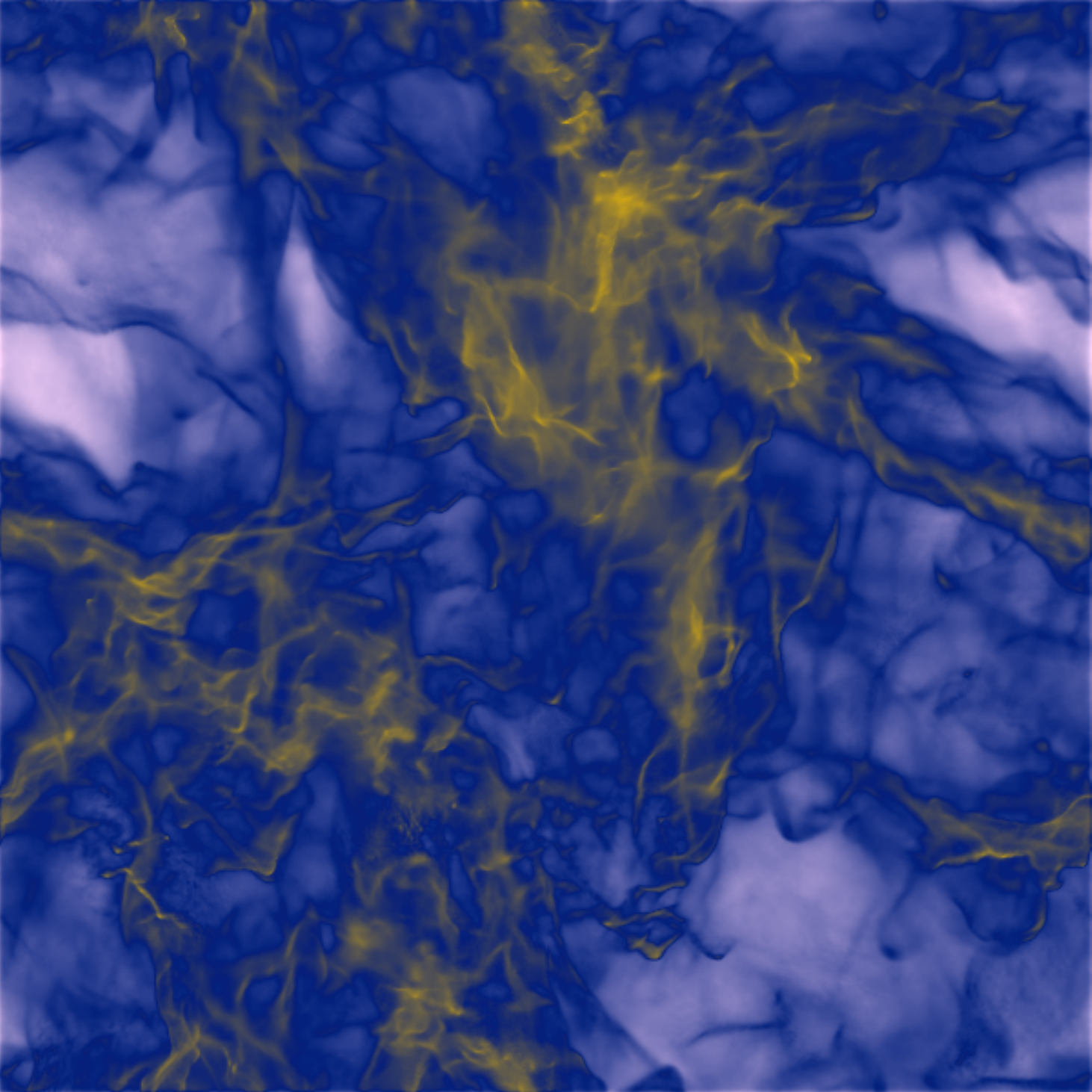} 
 &\includegraphics[width=0.28\linewidth]{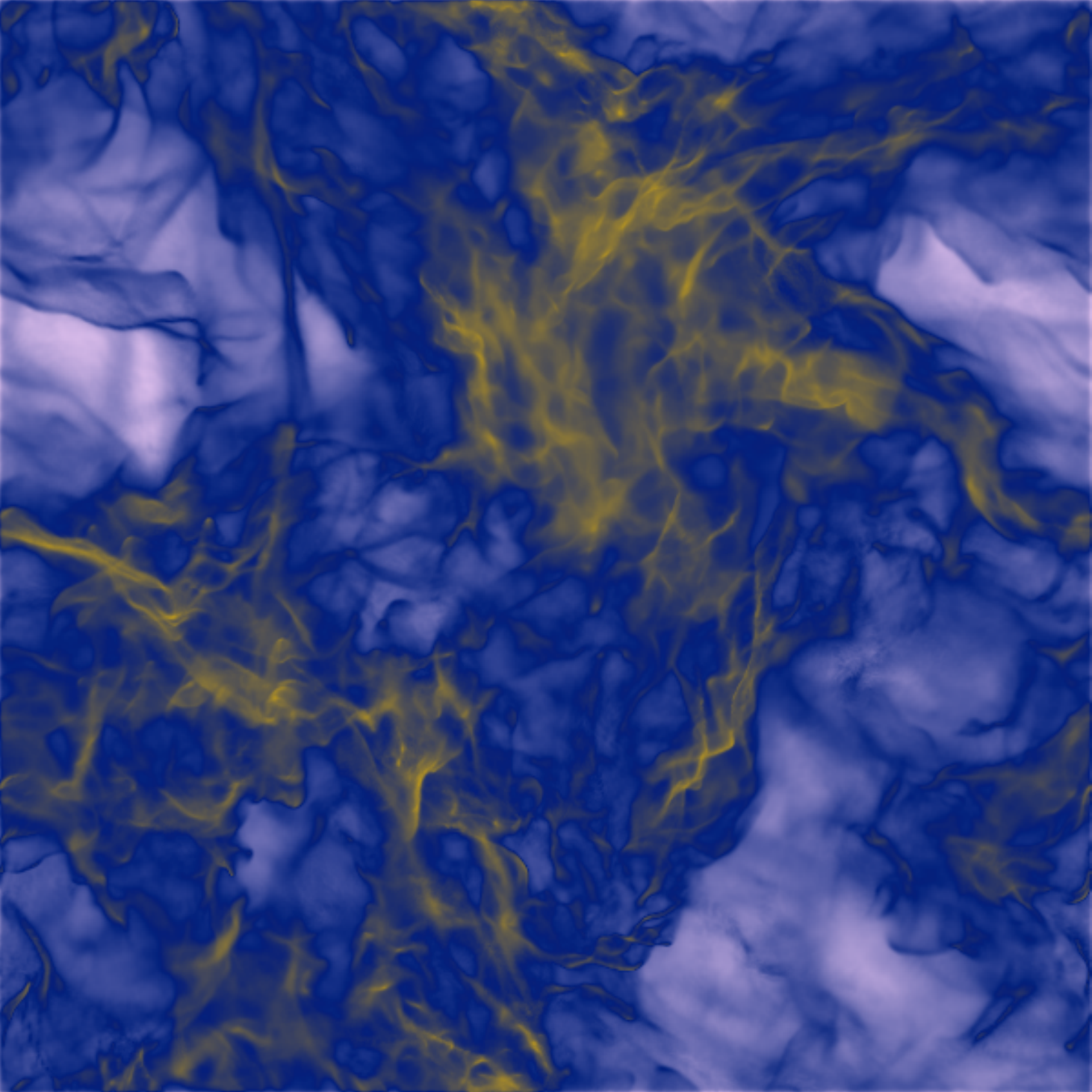} 
 & \includegraphics[height=0.28\linewidth]{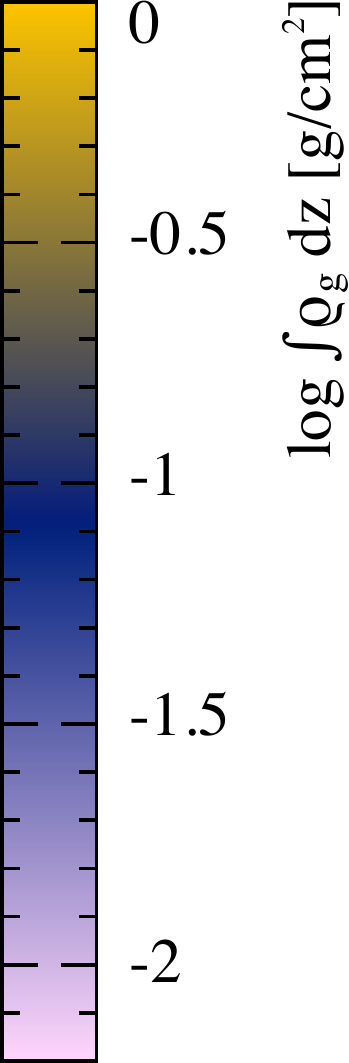} \\
\includegraphics[width=0.28\linewidth]{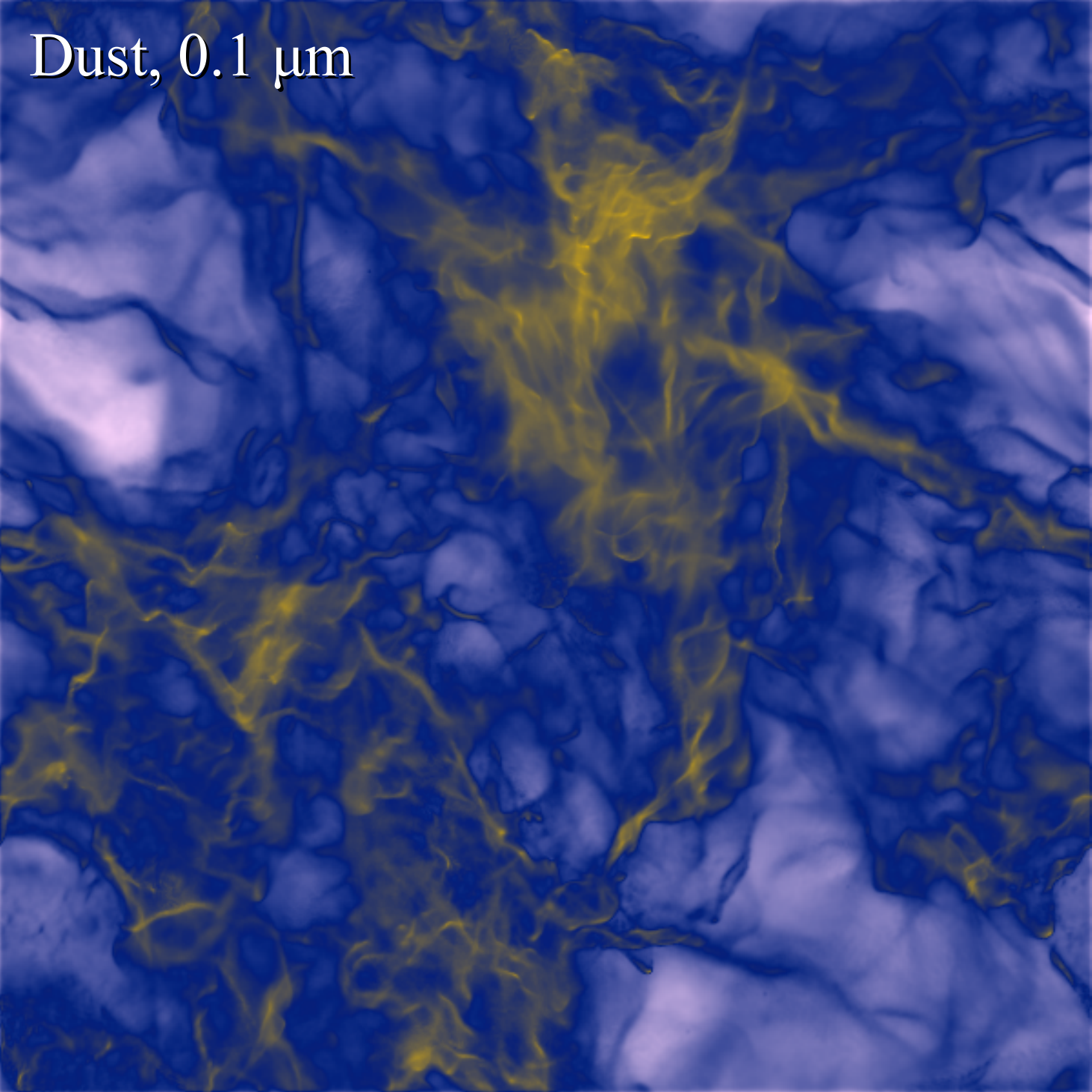} 
 & \includegraphics[width=0.28\linewidth]{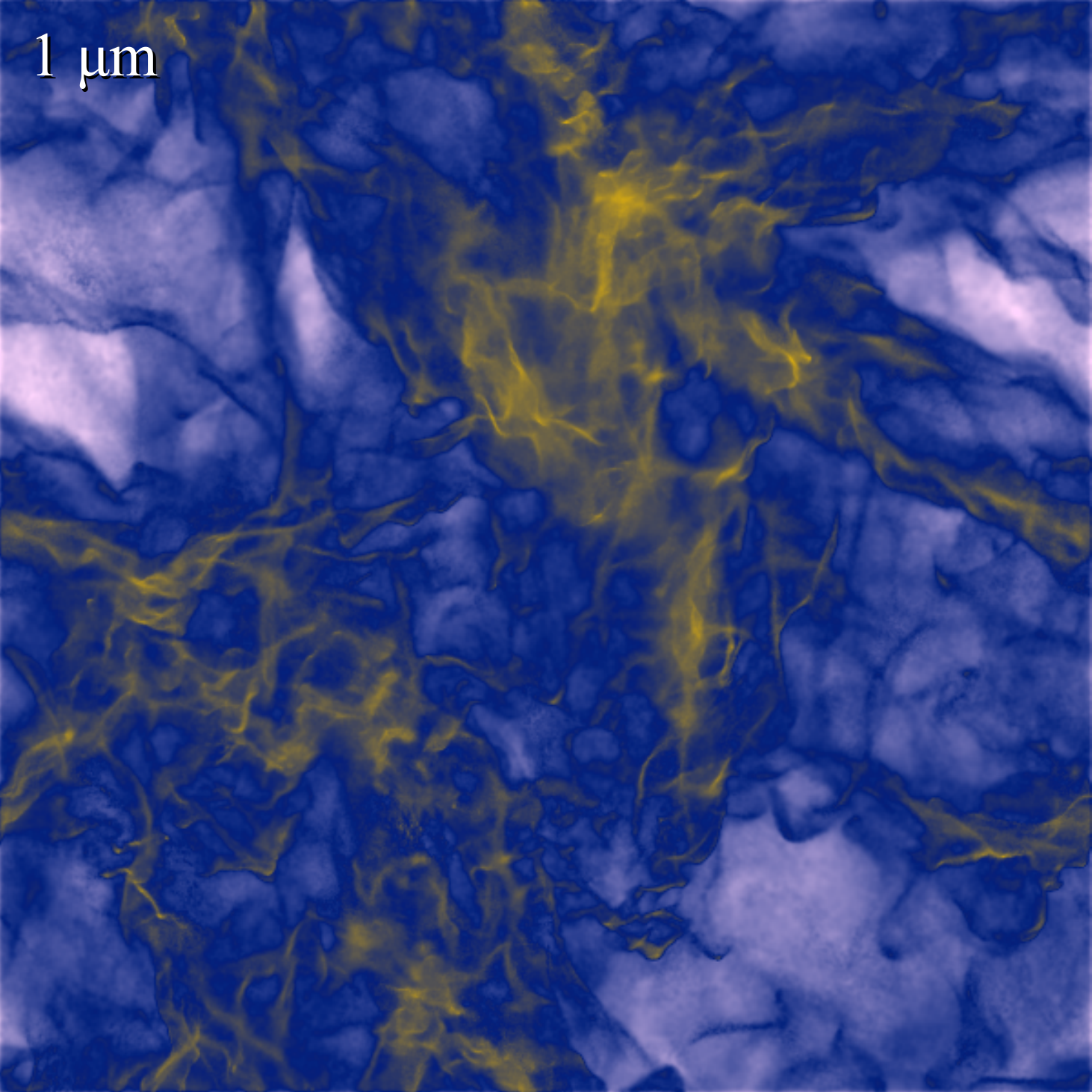} 
 & \includegraphics[width=0.28\linewidth]{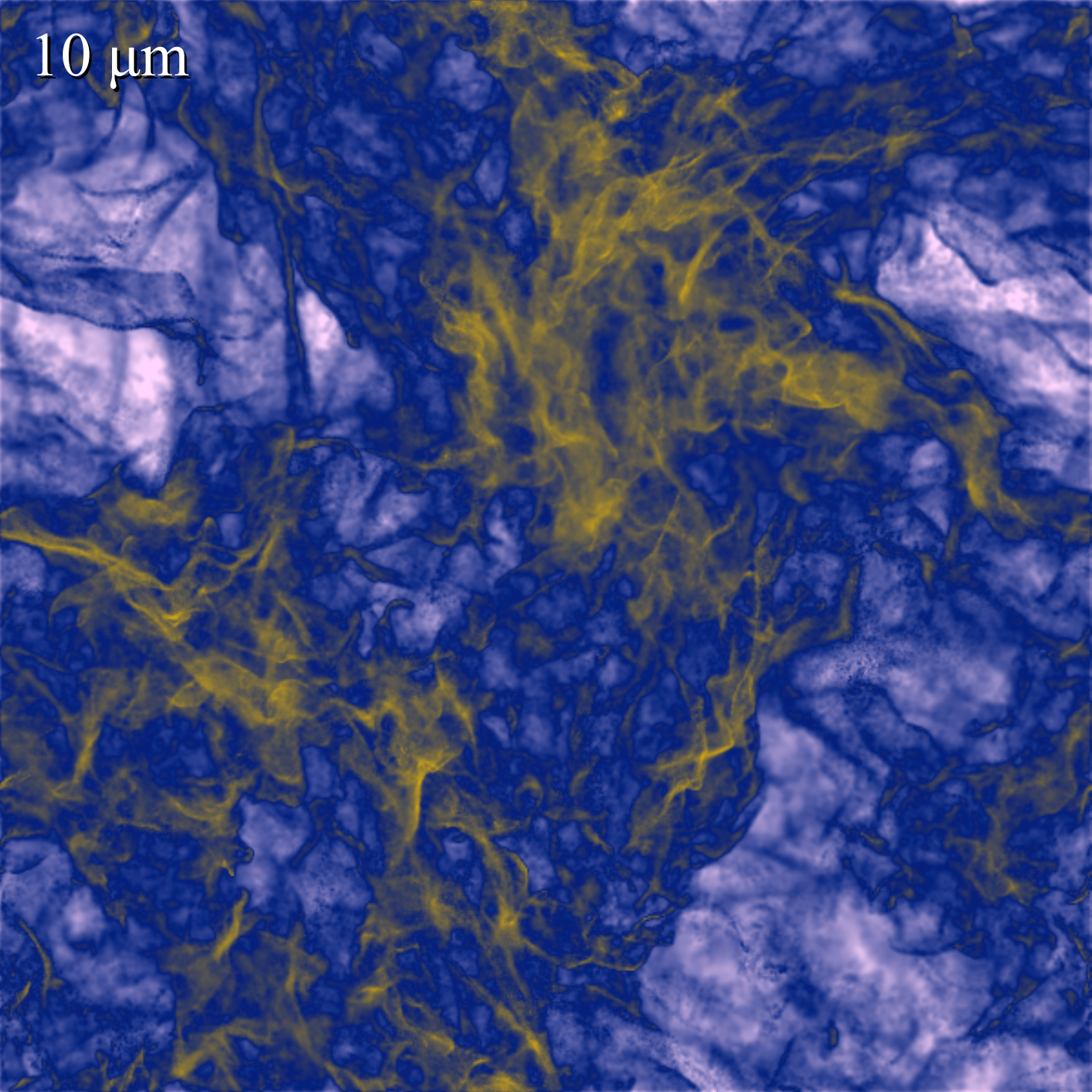} 
 & \includegraphics[height=0.28\linewidth]{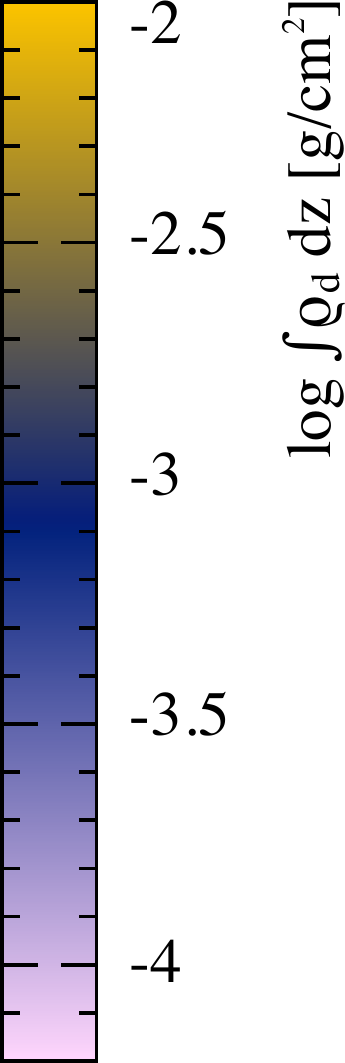} \\
\includegraphics[width=0.28\linewidth]{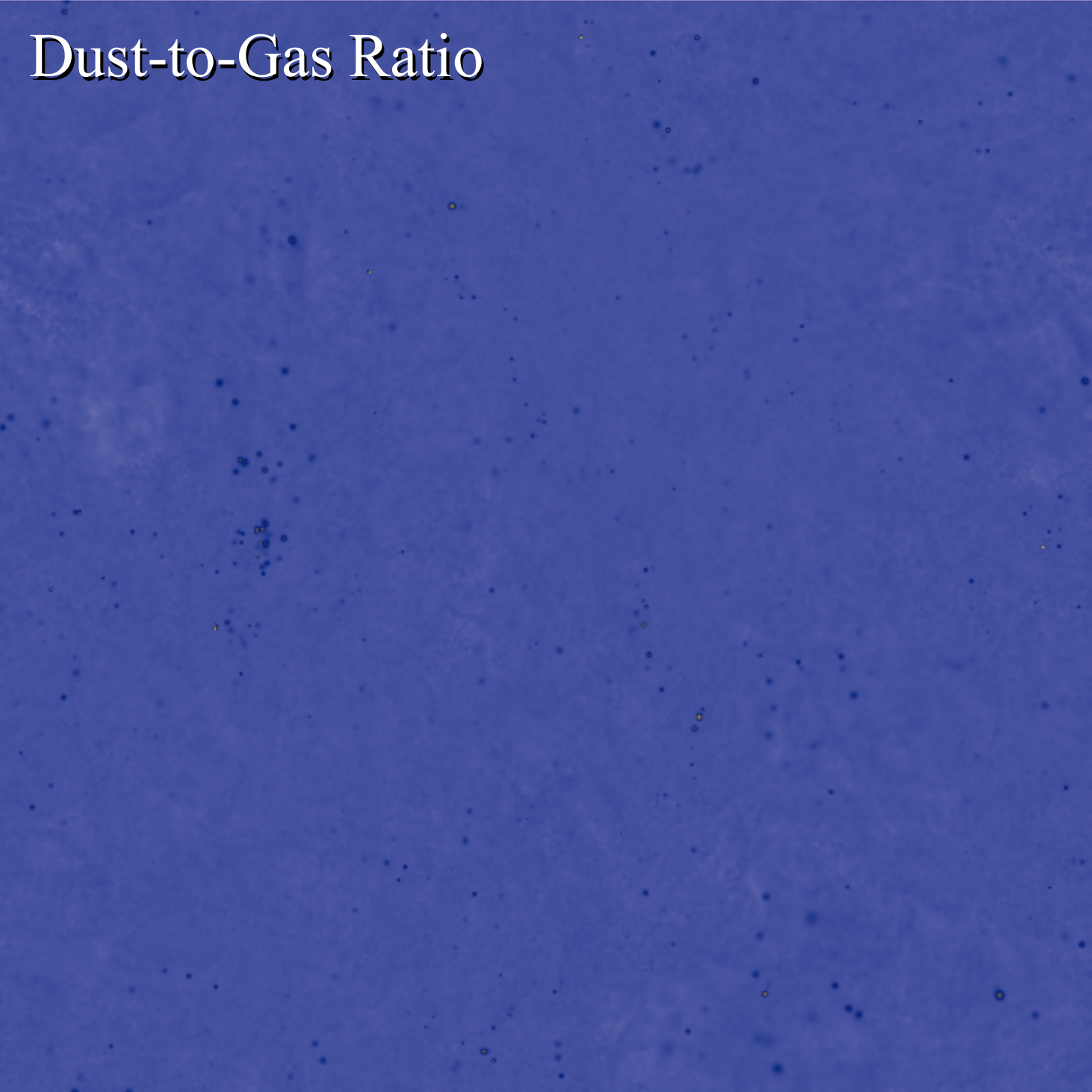} 
 & \includegraphics[width=0.28\linewidth]{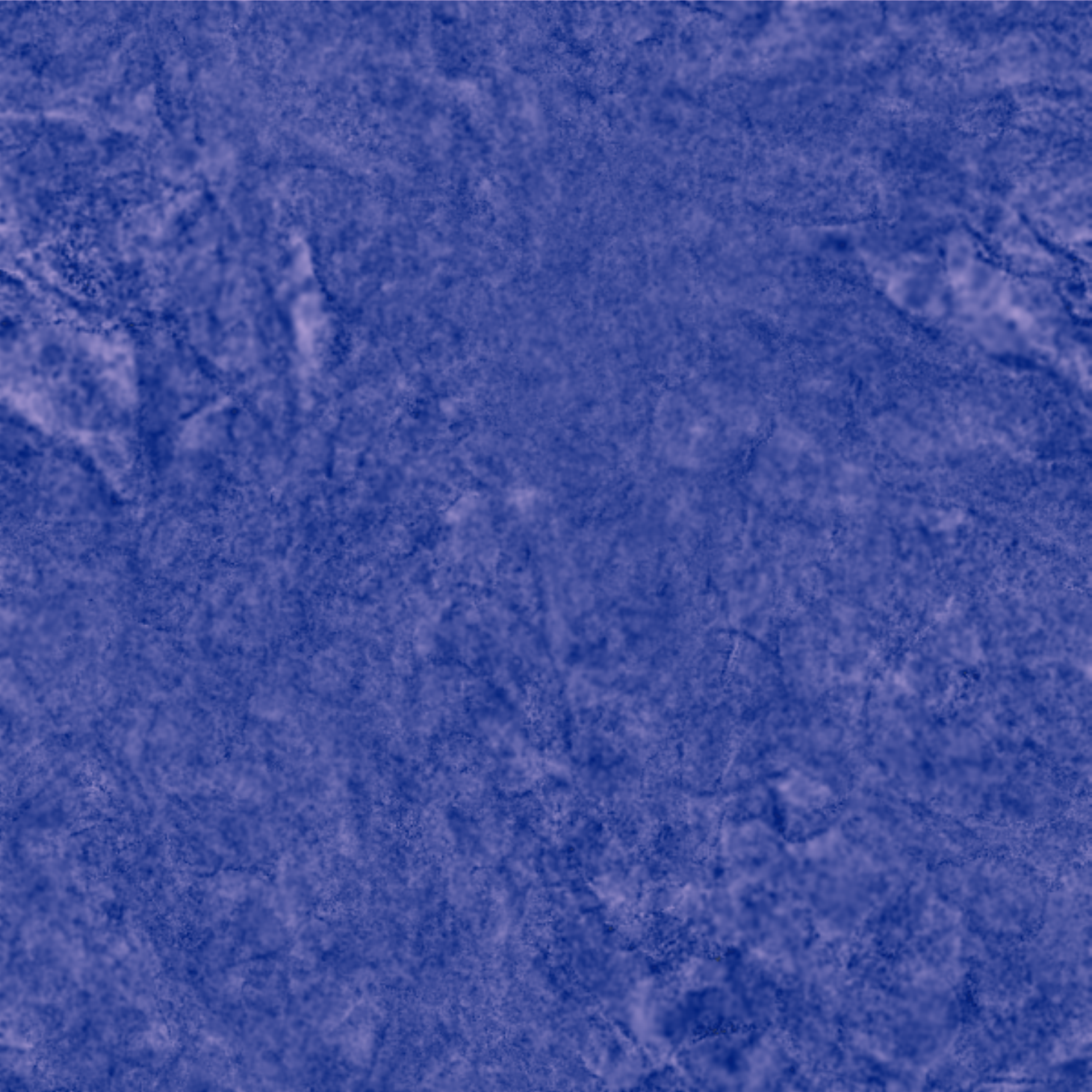} 
 & \includegraphics[width=0.28\linewidth]{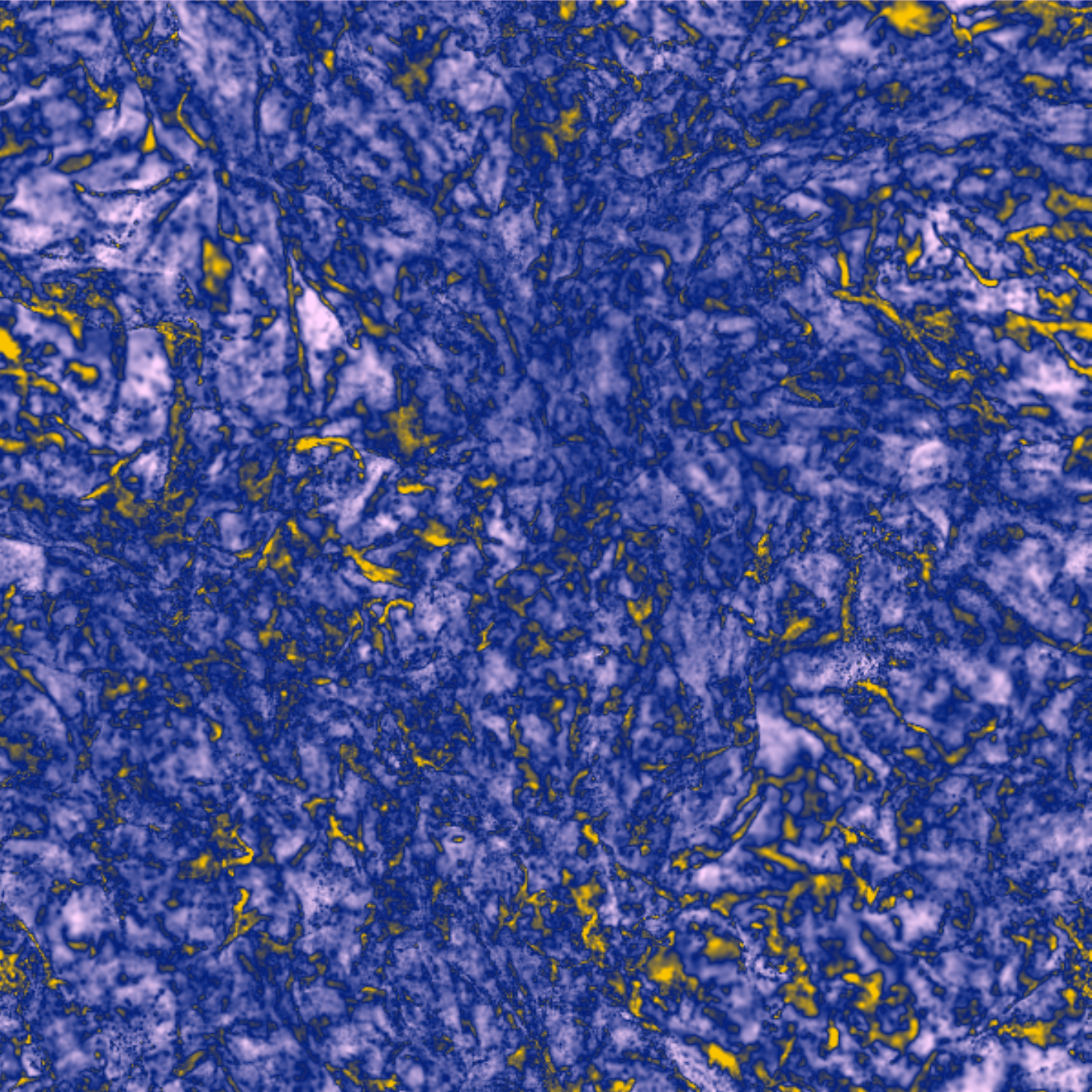} 
 & \includegraphics[height=0.28\linewidth]{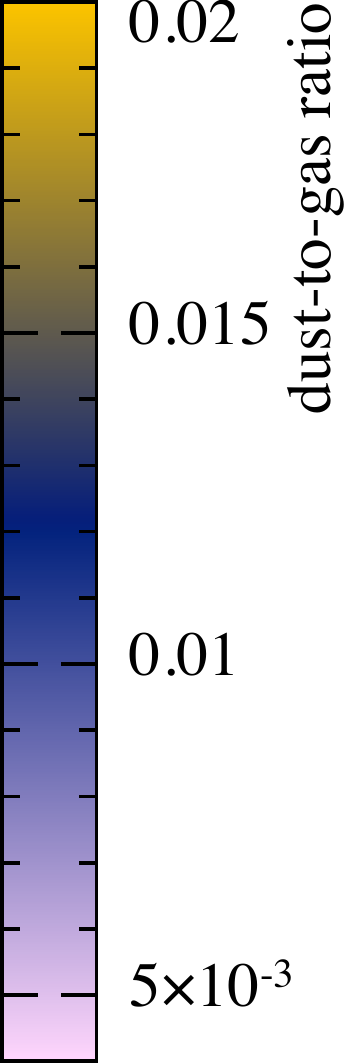} \\
\end{tabular}
\caption{Column density of gas (top row) and dust (centre row) and the dust-to-gas ratio (bottom row) for 0.1, 1 and 10~$\mu$m dust grains (left to right) at $t/t_{\rm c}=4$ ($\approx$2.93 Myr). The large scale structure of the dust traces the gas in all cases. For 0.1~$\mu$m grains there no discernible difference between the gas column density and the column dust density. Large, 10~$\mu$m grains (right), show a preferential concentration towards dense regions.}
\label{fig:col}
\end{figure*}
\subsection{Dust physics}

We model the dust/gas fluid mixture as a single fluid, with each element of fluid representing a combination of dust and gas \citep{lp14a, lp14b, pl15}. We solve the equations
\begin{align}
\frac{{\rm d}\rho}{{\rm d}t} &= - \rho (\nabla \cdot {\bm v}) , \label{eq:dust1} \\
\frac{{\rm d}{\bm v}}{{\rm d}t} &= - \frac{\nabla P_{\rm g}}{\rho} , \label{eq:dust2} \\
\frac{{\rm d}\epsilon}{{\rm d}t} &= - \frac{1}{\rho} \nabla \cdot \left( \epsilon t_{\rm s} \nabla P_{\rm g} \right) , \label{eq:dust4}
\end{align}
where ${\rm d}/{\rm d}t \equiv \partial / \partial t + {\bm v} \cdot \nabla$ is the material derivative, $P_{\rm g}$ is the thermodynamic gas pressure, $\rho$ is the sum of gas and dust densities, $\rho = \rho_{\rm g} + \rho_{\rm d}$, where subscripts represent the gas and dust, respectively, and $\epsilon \equiv  \rho_{\rm d}/{\rho}$ is the dust fraction. The mixture moves at the barycentric velocity
\begin{equation}
{\bm v}  = \frac{\rho_{\rm g} {\bm v}_{\rm g} + \rho_{\rm d}{\bm v}_{\rm d}}{\rho_{\rm g} + \rho_{\rm d}} .
\end{equation}
Gas and dust densities may be obtained from the total density and the dust fraction according to $\rho_{\rm g} = (1 - \epsilon) \rho$ and $\rho_{\rm d} = \epsilon \rho$. This means the dust-to-gas ratio may be expressed solely in terms of the dust fraction as
\begin{equation}
\frac{\rho_{\rm d}}{\rho_{\rm g}} = \frac{\epsilon}{1 - \epsilon} .
\end{equation}
Finally, we adopt an isothermal equation of state
\begin{equation}
P = c_{\rm s}^2 \rho_{\rm g} = c_{\rm s}^2 (1 - \epsilon) \rho,
\end{equation}
where the backreaction of the dust on the gas modifies the sound speed in the dust/gas mixture according to $\tilde{c}_{\rm s} = c_{\rm s} (1 + \rho_{\rm d} / \rho_{\rm g})^{-1/2}$.

Equations~\ref{eq:dust1}--\ref{eq:dust4} make use of the `terminal velocity approximation'. This is valid when the stopping time of dust grains is short compared to the dynamical time, occurring when the drag coefficient is large, i.e. when dust grains are small. We assume an Epstein drag prescription, appropriate for small grains. Assuming compact, spherical dust grains, the dust stopping time is 
\begin{equation}
t_{\rm s} = \frac{\rho_{\rm grain} s_{\rm grain}}{(\rho_{\rm d} + \rho_{\rm g}) c_{\rm s}} \sqrt{\frac{\pi \gamma}{8}} ,
\label{eq:ts}
\end{equation}
where $\rho_{\rm grain}$ is the intrinsic density of the dust grains, $s_{\rm grain}$ is the dust grain size, $c_{\rm s}$ is the speed of sound and $\gamma$ is the adiabatic index. Expressed in a manner appropriate for molecular clouds, this is
\begin{gather}
t_{\rm s} = 7.5 \times 10^{3}~{\rm yr} \left(\frac{\rho_\text{grain}}{3~\text{g cm}^{-3}} \right)  \left( \frac{s_\text{grain}}{0.1~\mu\text{m}} \right) \times \nonumber \\
\hspace{1cm} \left( \frac{c_{\rm s}}{0.2~\text{km s}^{-1}} \right)^{-1} \left( \frac{\rho}{10^{-20}~\text{g cm}^{-3}} \right)^{-1} .
\end{gather}
This timescale is shorter than the dynamical time for all grain sizes we consider, with the terminal velocity approximation becoming marginal only in the lowest density gas for our largest grain size (10 $\mu$m). Although the rms velocity of the turbulence is supersonic, the {\it relative} velocity between the gas and dust is not since grains experience strong drag. Therefore, there is no need to correct for relative supersonic motions, e.g.\ as in \citet{kwok75}.

\subsection{Numerical method}

We use the {\sc Phantom} smoothed particle hydrodynamics (SPH) code \citep{phantom}. Dust is modelled using the `one fluid' method of \citet{lp14a, lp14b} and \citet{pl15}, which is accurate and explicit for small dust grains (high drag) in the terminal velocity approximation. Our dust scheme exactly conserves gas, dust and total mass, along with linear momentum, angular momentum and energy to the accuracy of the timestepping. The scheme has been extensively benchmarked against the analytic solutions for linear waves and dusty shocks \citep{lp12a,lp14b}. Furthermore, both the one and two fluid dust algorithms in {\sc Phantom} have been previously used to simulate dust in protoplanetary discs \citep[e.g.][]{dipierroetal15,ragusaetal17}. We have also used {\sc Phantom} for previous studies of supersonic turbulence in both hydrodynamics and magnetohydrodynamics, including quantitative comparisons to results obtained with the grid-based code {\sc Flash} \citep*{pf10,tpf16}. \citet{phantom} gives full details of the dust-gas algorithm, turbulence driving routine, and SPH algorithms in {\sc Phantom}. This is the first application of our one fluid dust algorithm to molecular clouds.

\subsection{Initial conditions}

We assume a uniform, periodic box $x,y,z\in [0,L]$ with $L=3$ pc per side, adopting an isothermal sound speed $c_{\rm s}=0.2$ km s$^{-1}$ corresponding to a temperature of $\approx 11.5$ K. The mean total density (gas plus dust) is $\rho_0 = 10^{-20}$ g cm$^{-3}$. For these calculations, the maximum density produced by the turbulence is $\approx 10^{-17}$ g cm$^{-3}$, so it is reasonable to assume the gas remains isothermal. We neglect the self-gravity of the mixture. Turbulence is initiated and sustained at rms velocity Mach 10 ($\mathcal{M} = 10$), with a corresponding turbulent crossing time of $\tau=L/(2 \mathcal{M} c_{\rm s}) \approx 0.733$~Myr. Dusty shocks at this Mach number are expected to be of `J-type', with a sharp jump in the gas properties \citep{lw17}. Dust properties also undergo a sharp jump since the stopping length is short. We evolve the calculations for 20 dynamical times or about $14.66$ Myr. This may be longer than expected lifetimes for molecular clouds, but is necessary to ensure statistically meaningful results.

 We set the initial dust fraction assuming an initial dust-to-gas mass ratio of $1\%$ everywhere. We assume an intrinsic density of 3 g cm$^{-3}$ for the dust grains, representing a combination of carbonaceous (2.2 g cm$^{-3}$) and silicate grains (3.5 g cm$^{-3}$) \citep{draine03}. Simulations were performed with $0.1$, $1$ and $10$ $\mu$m sized grains, with a separate simulation for each grain size. 

\section{Results}
\label{sec:results}

\begin{figure*}
\centering
\setlength{\tabcolsep}{0.002\linewidth}
\begin{tabular}{cccl}
\includegraphics[width=0.28\linewidth]{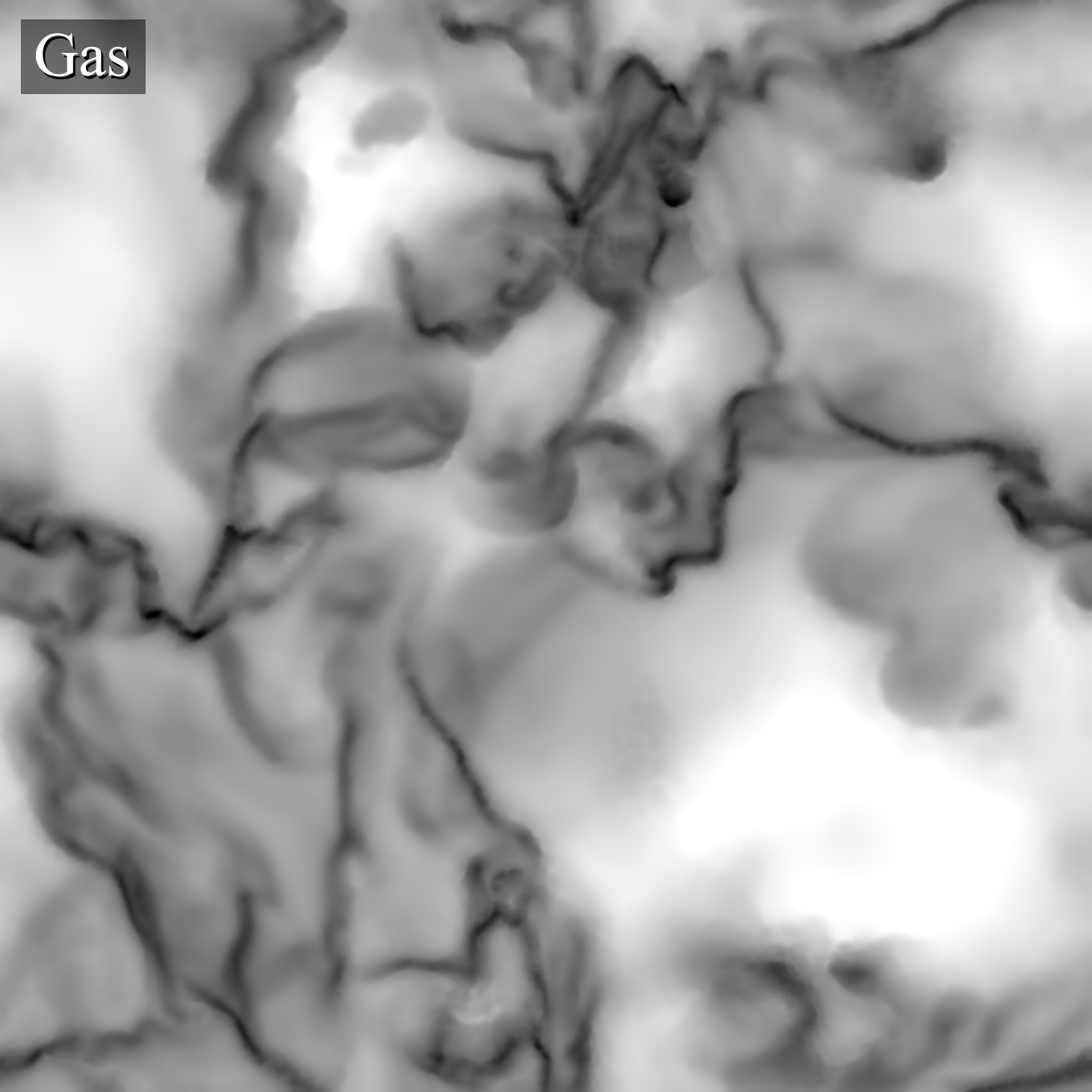} 
 & \includegraphics[width=0.28\linewidth]{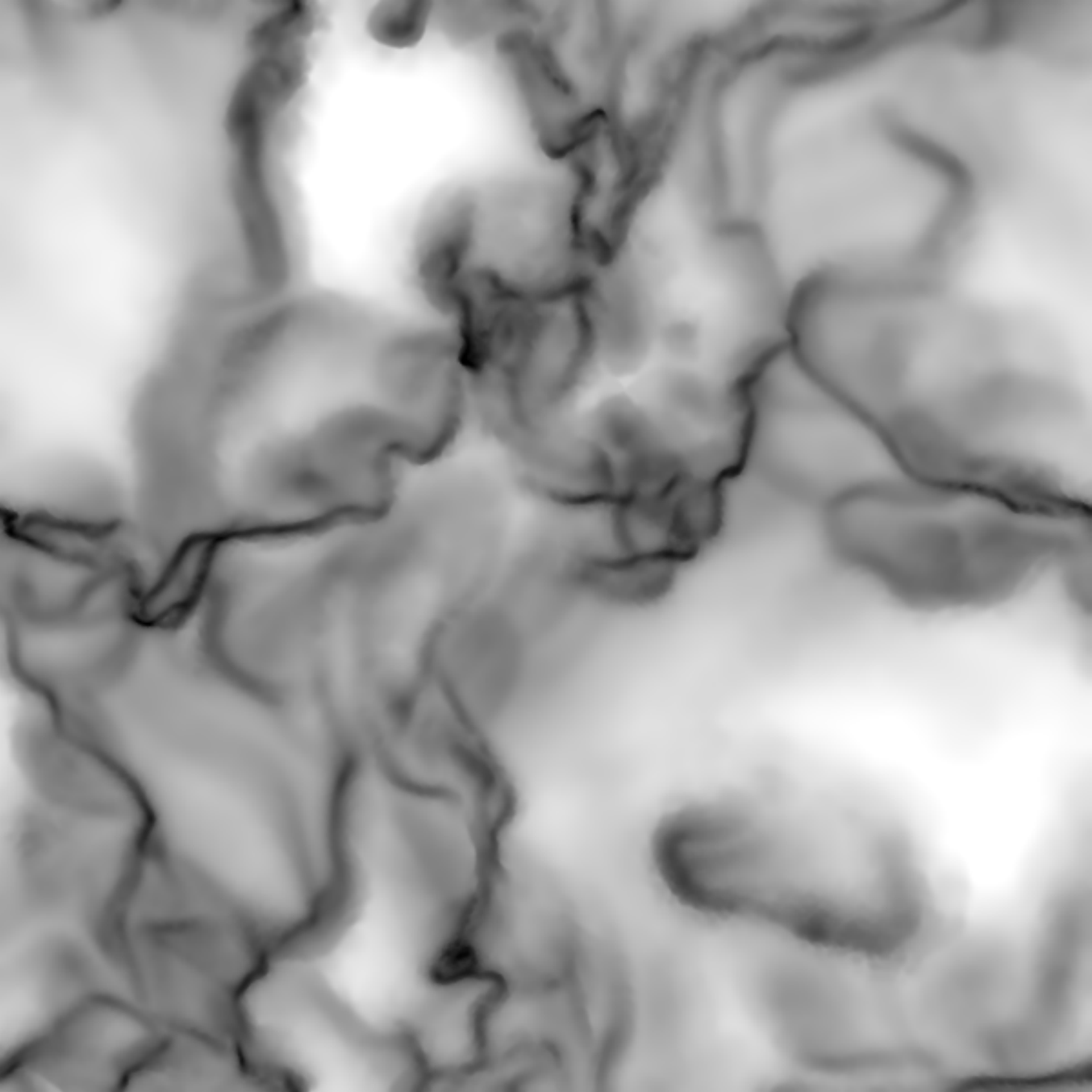} 
 & \includegraphics[width=0.28\linewidth]{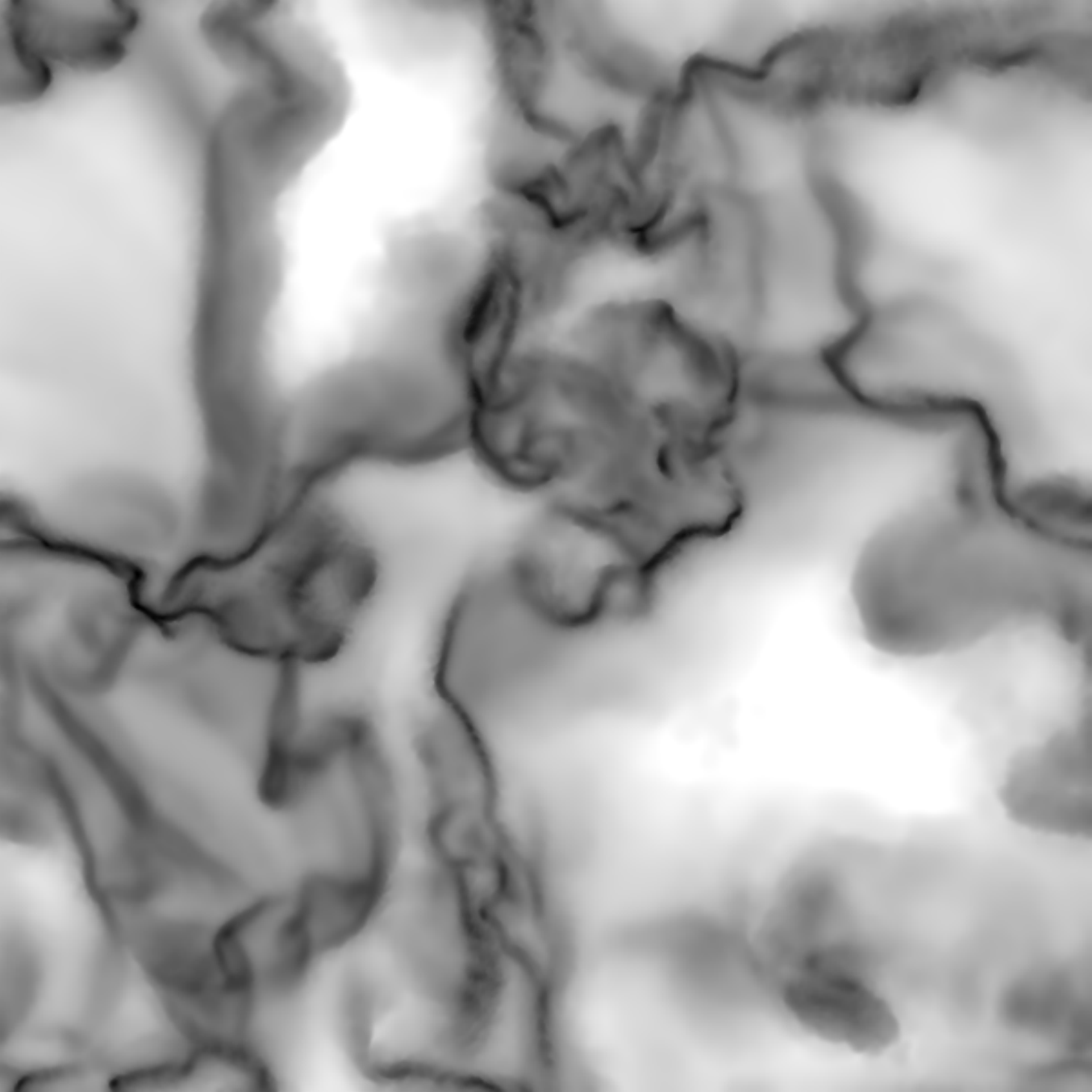} 
 & \includegraphics[height=0.28\linewidth]{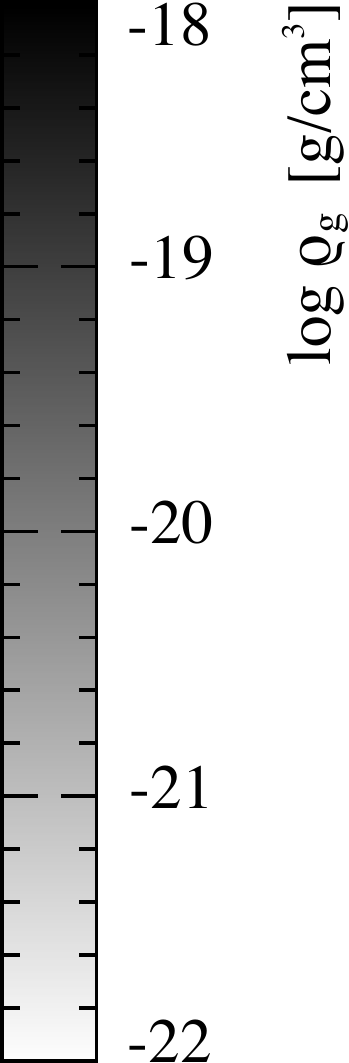}  \\
\includegraphics[width=0.28\linewidth]{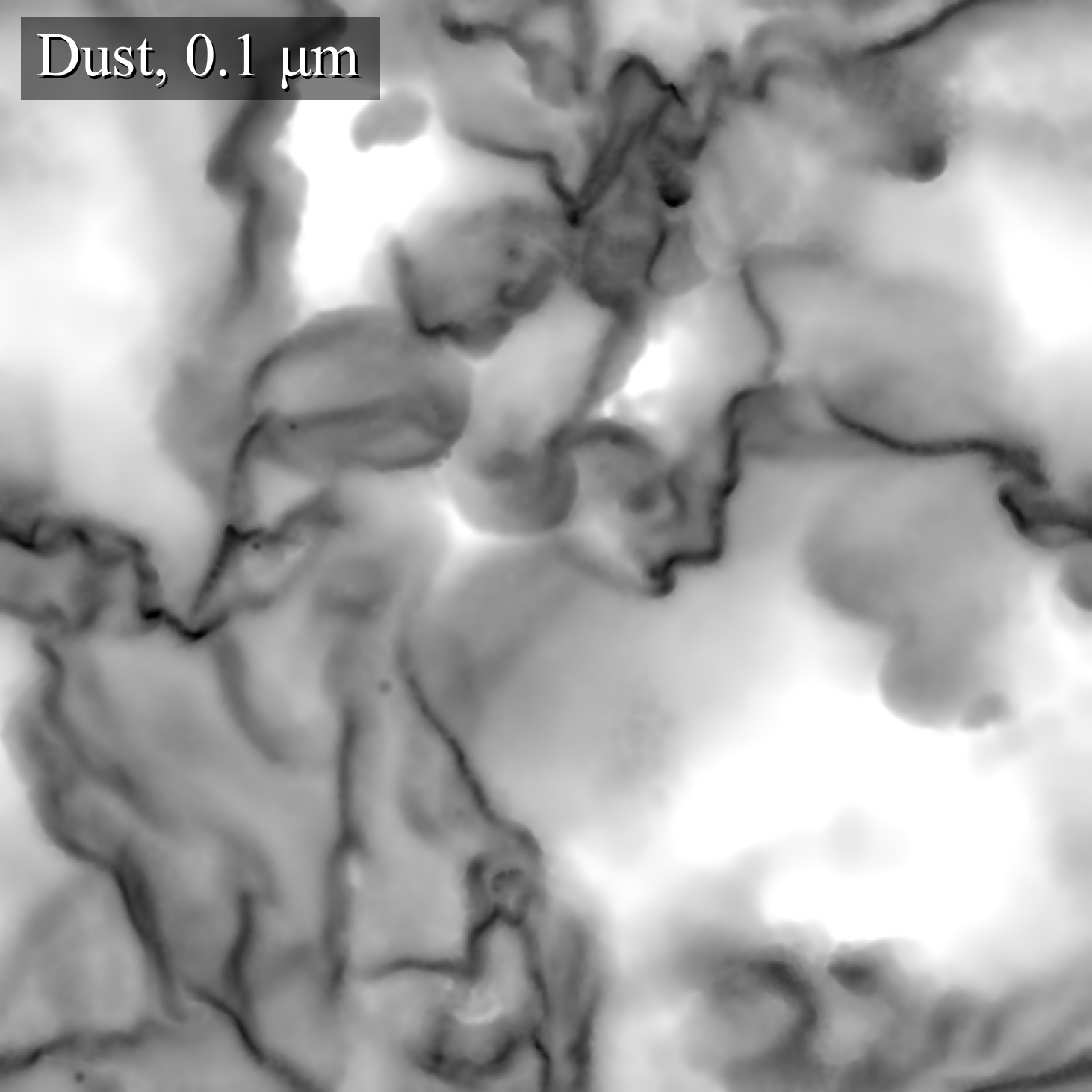} 
 & \includegraphics[width=0.28\linewidth]{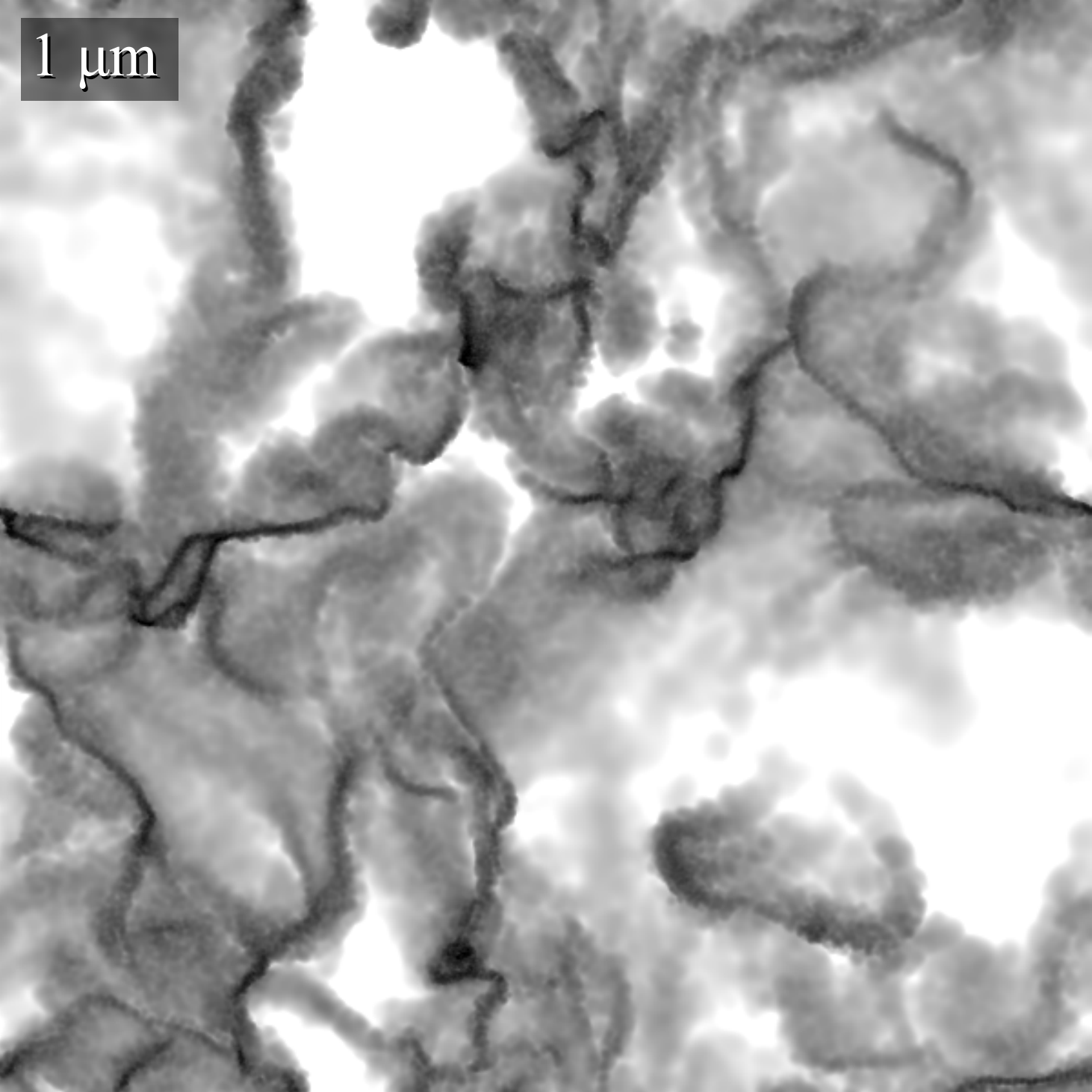} 
 & \includegraphics[width=0.28\linewidth]{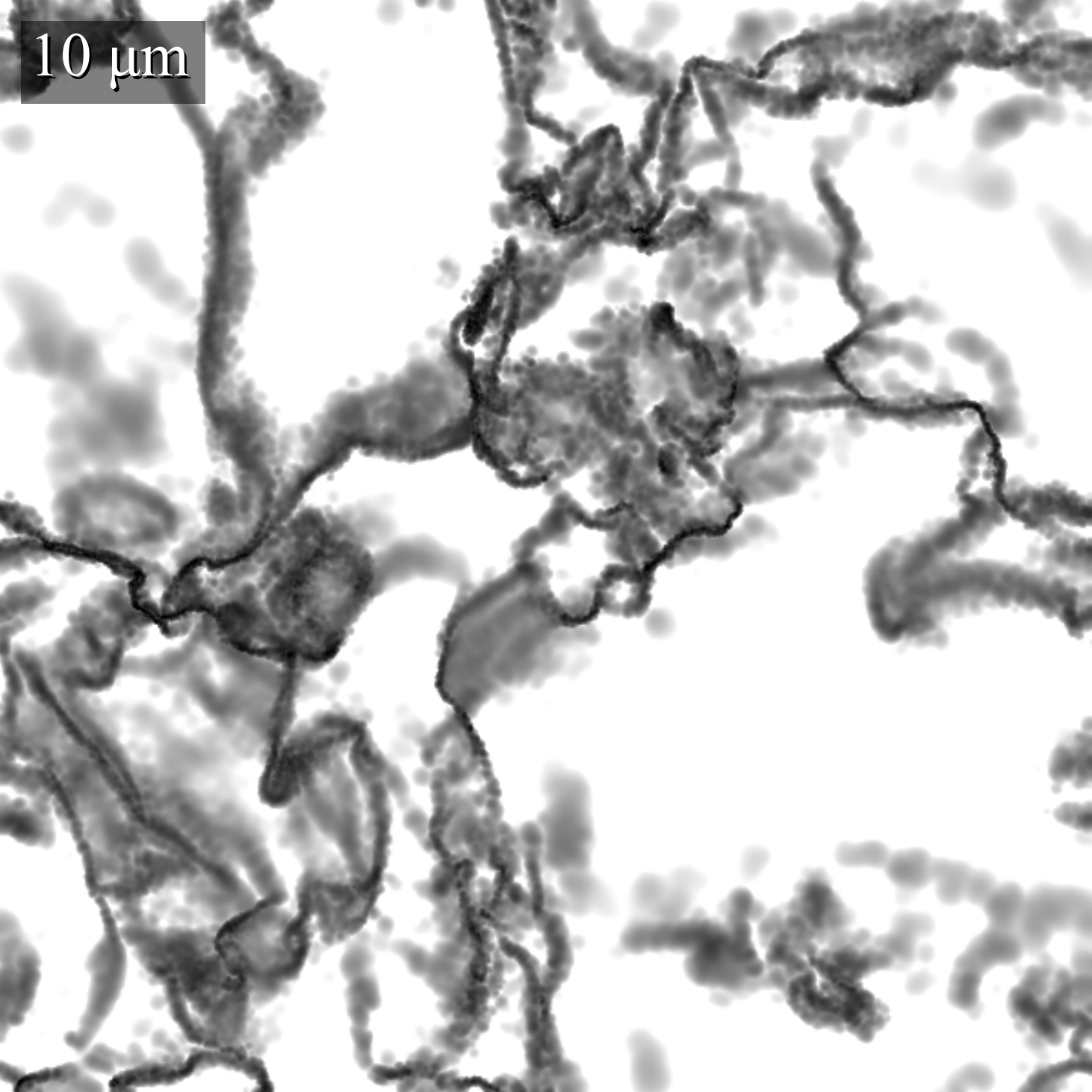} 
 & \includegraphics[height=0.28\linewidth]{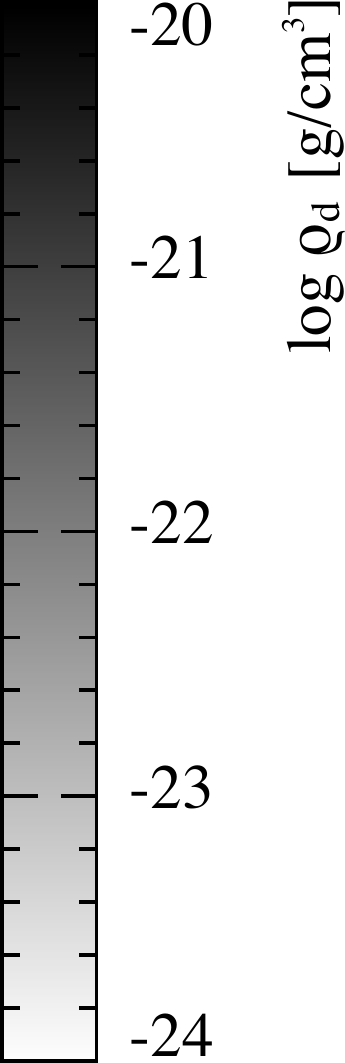} 
\end{tabular}
\caption{Turbulent `size-sorting' of dust grains. We show cross sections of the gas (top) and dust (bottom) density for 0.1, 1 and 10 $\mu$m dust grains (left to right) at $t/t_{\rm c}=4$ ($\approx$ 2.93 Myr). For small dust grains (0.1 $\mu$m), the dust almost perfectly traces the gas. As the dust grain size increases, the dust still traces the morphology of the gas filaments, but becomes preferentially concentrated in dense regions. }
\label{fig:slices}
\end{figure*}

\subsection{Column densities}

Figure~\ref{fig:col} compares the gas and dust column densities (top and centre rows, respectively) and the column dust-to-gas ratio (bottom row). For 0.1 $\mu$m dust grains (left column), the difference between column gas and dust density is imperceptible. For 1 $\mu$m grains, small differences are visible in the low density regions (middle column), but the overall morphology of the dust and gas column densities are similar. By contrast, dust column density for the large 10 $\mu$m grains shows distinct differences from the gas column density in both low and high density regions (right column), reflected in small-scale variations in the dust-to-gas ratio (bottom right panel). However, even for the largest grain size we simulated (10~$\mu$m), the morphology of the dust and gas column densities remain closely correlated, and dust column density remains an excellent tracer of the gas.

\subsection{Size sorting of dust grains}
Figure~\ref{fig:slices} shows cross sections of the gas and dust densities in the midplane of our computational domain. The gas density structure is similar between the 0.1, 1 and 10~$\mu$m dust grain calculations (top panels). For 0.1~$\mu$m grains (left column), the dust density closely matches the gas density, as reflected by the nearly uniform dust-to-gas ratio in Figure~\ref{fig:col}. For 1~$\mu$m grains (centre panels), low density regions appear diminished in dust compared to the gas. This effect is more pronounced for 10~$\mu$m grains (right panels). In this case, dust filaments remain correlated with gas filaments, but are thinner, with a sharper contrast between the low and high dust density regions (also seen in Figure~\ref{fig:col}), and with dust concentrated towards the dense gas filaments. The dust-to-gas ratio is increased by up to an order of magnitude within the filaments.

\begin{figure}
\centering
\includegraphics[width=\columnwidth]{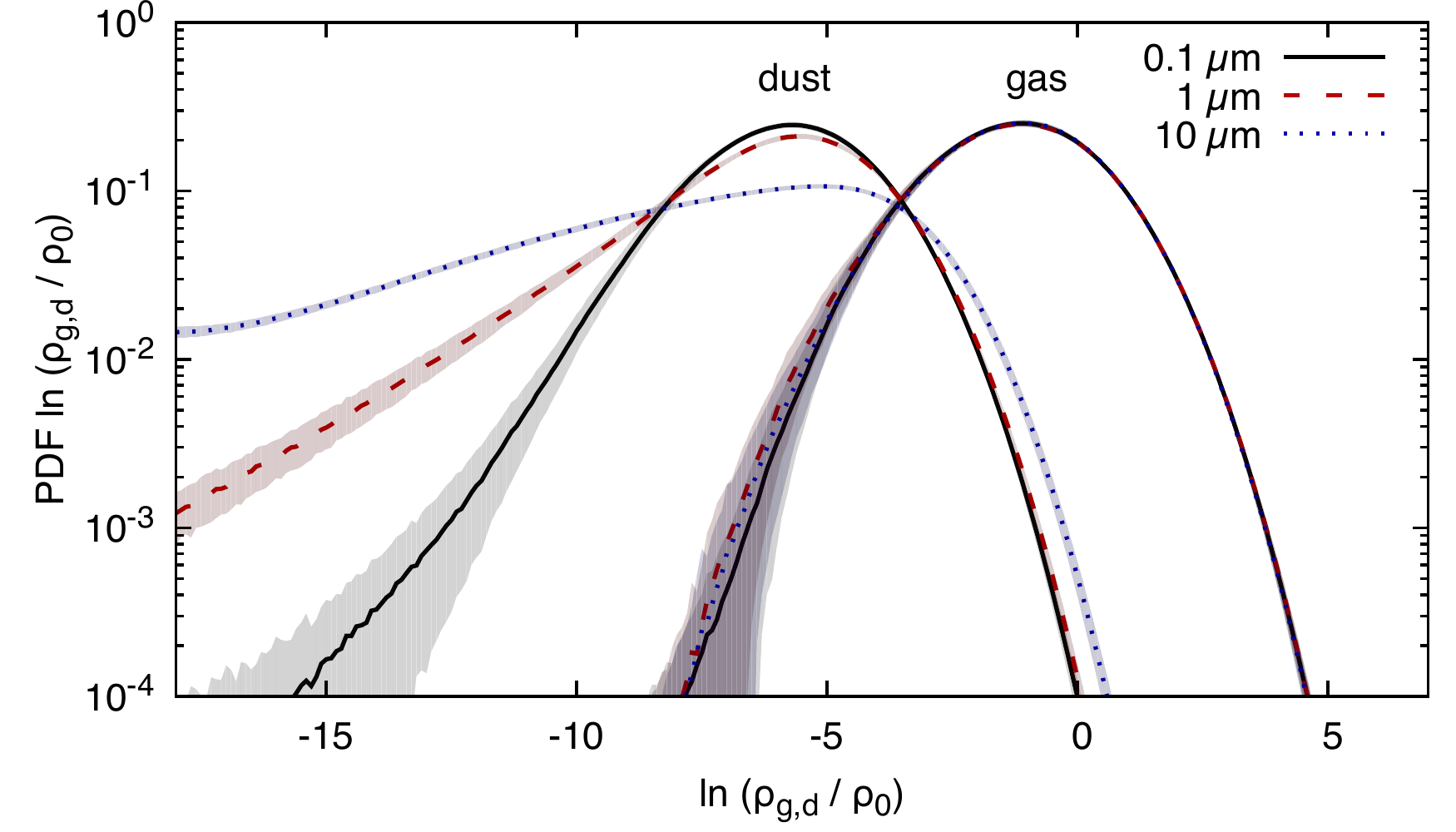}
\vspace{-0.4cm}
\caption{Time-averaged volume weighted PDFs of $\ln(\rho_{\rm g} / \rho_0)$ and $\ln(\rho_{\rm d} / \rho_0)$. Shaded regions represent the standard deviation of the time-averaging. The gas density is log-normal. Small 0.1 $\mu$m dust grains show a log-normal PDF, mirroring the gas. Larger grains show a skewed distribution, with regions of both low and high dust densities being more common, indicating the preferential concentration of large grains into dense regions.}
\label{fig:lnrhogdpdf}
\end{figure}

Figure~\ref{fig:lnrhogdpdf} quantifies these differences, showing the Probability Density Functions (PDFs) of the gas and dust densities. The gas density PDF is log-normal, characteristic of supersonic, isothermal turbulence \citep[e.g.][]{vs94, pvs98}, and is similar for all three grain sizes, indicating that the gas density PDF is not significantly affected by the backreaction of the grains on the gas.

\begin{figure}
\centering
\includegraphics[width=\columnwidth]{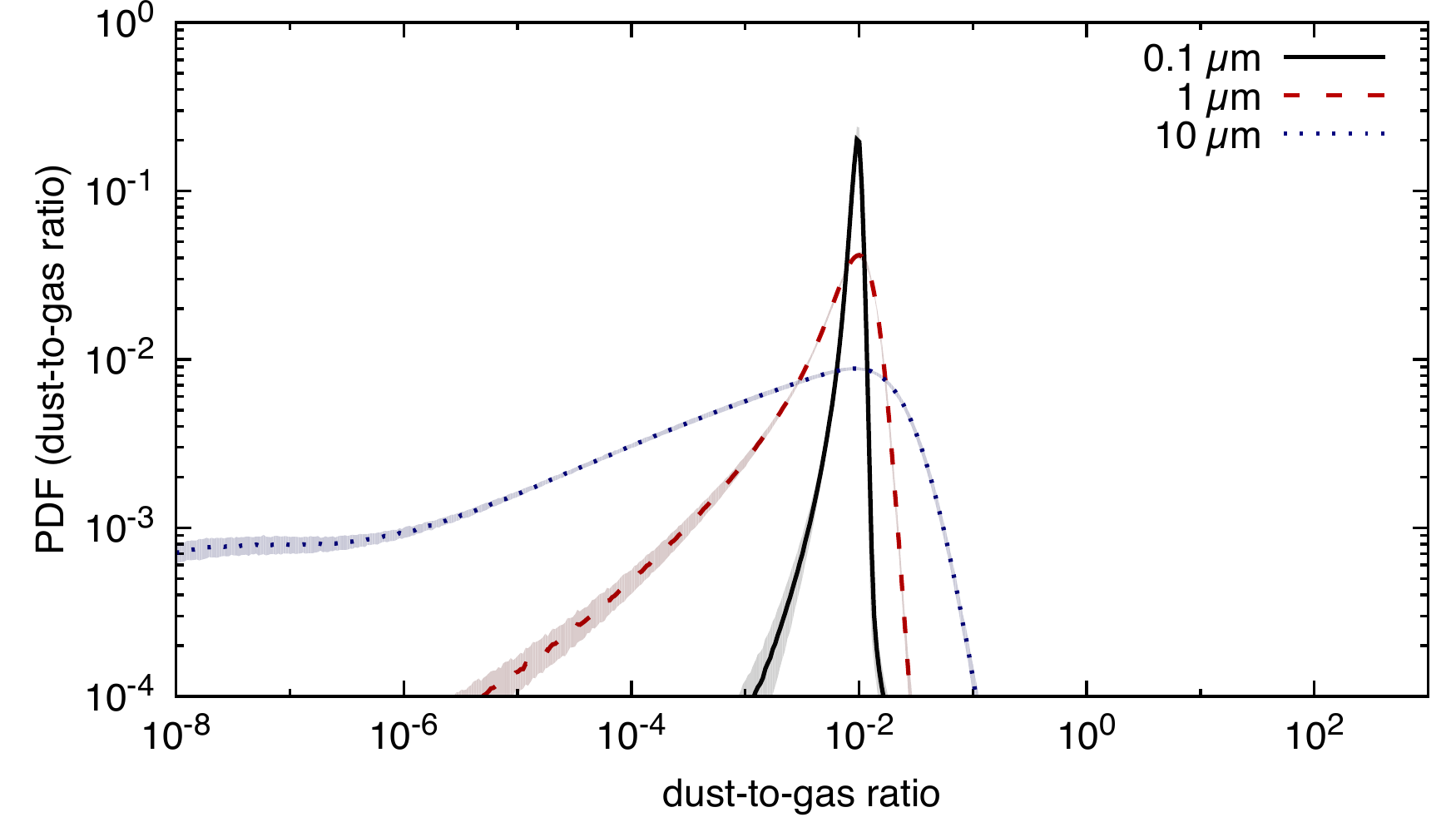} 
\vspace{-0.4cm}
\caption{Time-averaged volume weighted PDFs of the dust-to-gas ratio. Shaded regions represent the standard deviation from time-averaging. The dust-to-gas ratio for all grain sizes is peaked at $1\%$. Minimal variation in the dust-to-gas ratio occurs for $0.1$ $\mu$m grains, but the PDFs broaden with increasing grain size caused by `size-sorting' of large grains, meaning larger volumes of the cloud are either dust enriched or dust depleted.}
\label{fig:dgrpdfs}
\end{figure}
 
The dust density PDF --- shifted towards lower densities by the mean dust-to-gas ratio --- is also log-normal for 0.1~$\mu$m dust grains. This is expected since the dust remains tightly coupled to the gas (see Figures~\ref{fig:col} \& \ref{fig:slices}). For 1 and 10~$\mu$m grains, the high density tail of the dust PDF remains log-normal, matching the gas density. However, the low density tail broadens as the grain size increases, due to the dependence of $t_{\rm s}$ on gas density. The stopping time increases in low density gas, allowing large grains to decouple, but decreases within dense filaments, trapping dust. This leads to transfer of large grains from low density gas into filaments (`size-sorting'), causing the broadening of the PDF seen in Figure~\ref{fig:lnrhogdpdf}.

\subsection{Variations in the dust-to-gas ratio}

Figure~\ref{fig:dgrpdfs} shows PDFs of the dust-to-gas ratio. For 0.1~$\mu$m grains, the dust-to-gas ratio is sharply peaked at 1\%. The maximum in the PDF for 1 and 10~$\mu$m dust grains remains close to 1\%, but with a modest increase in higher dust-to-gas ratios and a broad distribution of low dust-to-gas ratios. This occurs due to the `size-sorting' of dust grains discussed previously.

Table~\ref{tbl:dgrs} quantifies the volumetric mean dust-to-gas ratios in all of our calculations, with deviations reflecting the 68th percentile about the median. For 0.1 $\mu$m dust grains, the mean is 0.92\%, close to the starting value of 1\%. The mean decreases with increasing grain size, dropping to 0.77\% for 1~$\mu$m grains and 0.56\% for 10~$\mu$m grains. However, the deviation increases, reflecting the broadened dust-to-gas PDF (Figure~\ref{fig:dgrpdfs}). The $128^3$ and $256^3$ particle calculations converge in mean dust-to-gas ratio, but the higher resolution calculations show an increased deviation. This is mainly due to these calculations sampling a broader range of gas densities, particularly at low densities, in which large grains preferentially decouple.

\begin{table}
\centering
\caption{Mean dust-to-gas ratio as a function of grain size and resolution.}
\begin{tabular}{ccSc}
\hline
Grain size & Resolution & Mean ($\times 10^{-2}$) \\
\hline
0.1 $\mu$m & $64^3$,$128^3$, $256^3$ & $0.95^{+0.0}_{-0.05}$, $0.93^{+0.05}_{-0.08}$, $0.92^{+0.08}_{-0.13}$ \\
1 $\mu$m & $64^3$,  $128^3$, $256^3$ & $0.86^{+0.18}_{-0.22}$, $0.79^{+0.31}_{-0.37}$, $0.77^{+0.41}_{-0.50}$ \\
10 $\mu$m & $64^3$,  $128^3$, $256^3$ & $0.63^{+0.47}_{-0.51}$, $0.55^{+0.57}_{-0.54}$, $0.56^{+0.72}_{-0.55}$ \\
\hline
\end{tabular}
\label{tbl:dgrs}
\end{table}

\section{Discussion}
\label{sec:discussion}

In our simulations, 0.1~$\mu$m dust grains remain well-coupled to the gas. This is not true for large grains ($\gtrsim 10$ $\mu$m), a consequence of the dust stopping time (Equation~\ref{eq:ts}) being proportional to grain size but inversely proportional to density. Though large grains remain coupled to the gas in dense regions, they can dynamically decouple in low density regions. This leads to the preferential concentration of large grains in dense filaments.

This `size-sorting' of grains may help to explain the different extinction laws observed along lines of sight that pass through molecular clouds in the Milky Way. Size-sorting leads to a preferential increase in the mean grain size in dense regions, potentially explaining extinction laws with $R_{\rm V}\gtrsim 5$ \citep[e.g.][]{wd01} without grain growth. Size-sorting also occurs in protoplanetary discs \citep{dd04, pinteetal07, pignataleetal17} and during protostellar collapse \citep{bla17}. Producing extinction maps from our calculations is complicated by the slightly different turbulence patterns induced in the gas by the backreaction of each grain size, meaning that direct stacking of the dust maps is not possible. What is needed are calculations that evolve multiple grain sizes simultaneously \citep{lp14c}, which we intend to address in a subsequent paper.
 
We find typical fluctuations in the dust-to-gas ratio for 0.1 $\mu$m grains of around 10\%, much smaller than the orders-of-magnitude fluctuations found by \citet{hl16} and \citet{lhs17}. Our one fluid dust model is specifically designed to be accurate at high drag, implying that such large fluctuations are a numerical artefact of using tracer particles to simulate dust \citep[c.f.][]{pf10, lp12a}. Indeed, recent calculations by \citet{bla17} found dust grains $\lesssim 10\mu$m in size closely follow the gas during the early stages of gravitational collapse in a molecular cloud core ($\rho_{\rm g} \approx 10^{-18}$ to $10^{-12}$ g cm$^{-3}$).

The main caveat is that we have neglected magnetic fields, which are important both for the dynamics of turbulence in molecular clouds (e.g. \citealt{molinaetal12}) and for the dust dynamics \citep{yld04}. Neither do we account for interstellar radiation, which may also affect the grain dynamics \citep{wb02}.

\section{Conclusions}
\label{sec:summary}

Does supersonic turbulence affect the dust-to-gas ratio in molecular clouds? It depends on the grain size. Our main conclusions are:
\begin{enumerate}
\item We find evidence for turbulent `size-sorting' of dust grains, whereby dynamical effects lead to the preferential concentration of large ($\gtrsim 10$ $\mu$m) grains into dense gas filaments.

\item Local fluctuations in the dust-to-gas ratio around the mean were $\approx$10\% for 0.1 $\mu$m and $\approx$40\% for 10 $\mu$m grains. Larger grains ($\gtrsim 10$ $\mu$m) allow for larger variations, with maximum local dust-to-gas ratios increased by an order of magnitude in dense filaments. The large scale dust column density remains well correlated with the gas column density for all grain sizes.

\item Contrary to \citet{hl16}, we find that supersonic turbulence cannot introduce orders of magnitude fluctuations in the dust-to-gas ratio for $0.1$ $\mu$m grains. We find no evidence for `totally metal' star forming cores \citep{hopkins14}.
\end{enumerate}

\vspace{-0.7cm}
\section{Acknowledgments}
We thank Christophe Pinte, Matthew Bate and Stella Offner for useful discussions. TST is supported by a CITA Postdoctoral Fellowship. DJP acknowledges Australian Research Council  grants FT130100034 and DP130102078. We used {\sc splash} \citep{splash}. We thank the referee, Jacco van Loon, for insightful comments.

\vspace{-0.6cm}
\bibliographystyle{mnras}
\bibliography{bib}

\label{lastpage}
\end{document}